\documentclass{PoS}
\usepackage{latexsym,amsmath,amstext}
\usepackage{mathrsfs}
\usepackage[utf8x]{inputenc}
\usepackage{epsfig}
\usepackage{float}
\usepackage{amssymb,fontenc,times,mathptmx,graphicx}
\newcommand\bef{\begin{figure}}
\newcommand\eef[1]{\label{fg:#1}\end{figure}}
\newcommand\beq{\begin{equation}}
\newcommand\eeq[1]{\label{#1}\end{equation}}
\newcommand\beqa{\begin{eqnarray}}
\newcommand\eeqa[1]{\label{#1}\end{eqnarray}}
\newcommand\bet{\begin{table}}
\newcommand\eet[1]{\label{tb:#1}\end{table}}

\newcommand\fgn[1]{Figure \ref{fg:#1}}

\newcommand\tbn[1]{Table \ref{tb:#1}}

\newcommand\etal{{\sl et al.\/}}
\newcommand\jhep{{\sl J.\ H.\ E.\ P.\/}\ }
\newcommand\np{{\sl Nucl.\ Phys.\/}\ }

\newcommand\prd{{\sl Phys.\ Rev.\ D\/}\ }
\newcommand\prl{{\sl Phys.\ Rev.\ Lett.\/}\ }

\newcommand\plt{{\sl Phys.\ Lett.\/}\ }

\newcommand{\IR}{{\scriptscriptstyle IR}}
\newcommand{\UV}{{\scriptscriptstyle UV}}

\title{UV Suppression by Smearing and Screening Correlators}
\author{Sourendu Gupta and \speaker{Nikhil Karthik} \\
        Department of Theoretical Physics,\\
   Tata Institute of Fundamental Research,\\
   Homi Bhabha Road, Mumbai 400005, India \\
\email{sgupta@theory.tifr.res.in}, \email{nikhil@theory.tifr.res.in}}

\abstract{ We investigate the mechanism of smearing in the APE, Stout,
HYP and HEX schemes through their effect on glue and quark Fourier
modes. Using this, we non-perturbatively tune the smearing parameters
to their optimum values. Smearing causes a super-linear improvement in
taste symmetry breaking in the high temperature phase of QCD. We use
optimal smearing in the high temperature phase and find close agreement
of meson screening masses with weak coupling predictions.  }

\FullConference{31st International Symposium on Lattice Field Theory LATTICE 2013\\
                 July 29 – August 3, 2013\\
                 Mainz, Germany}
\ShortTitle{UV suppression by smearing and screening correlators}
\begin{document}

\section{Introduction}
Smeared gauge links improve the scaling behaviour of staggered quarks.
It is believed that smearing achieves this by suppressing the dependence
of operators on high-momentum field modes. Also, hypercubic gauge link
smearing is gaining interest with the advent of HEX \cite{HEX} as it is
differentiable and can be implemented in dynamical simulations. Hence
it is important to check if gauge link smearing suppresses only the
ultraviolet modes while leaving the long distant physics intact. In this
presentation, we present our study on the effect of gauge link smearing
on the infrared and ultraviolet modes in both gluonic and quark sectors.

From \cite{saumen}, comparison of hadron screening masses at $1.5T_c$
from various quenched calculations using different valence quarks shows
complete disagreement between staggered quark results and results from
Clover and Overlap fermion calculations. In particular, the staggered
pseudoscalar/scalar screening masses lie anomalously below the lattice
free field theory. We were motivated to see if this discrepancy is due
to taste splitting in staggered fermion formulation.

The results presented here are based on \cite{nik}.

\section{Study of smearing}
We examined four schemes which are currently popular: APE \cite{APE},
HYP \cite{HYP}, Stout \cite{ST}, and HEX \cite{HEX}.  Each of these
smearing schemes have free parameter $\epsilon$ which determines
how much importance is given to link neighbours.  The APE and Stout
schemes have a single fattening parameter $\epsilon$, while HYP and HEX
schemes have three different $\epsilon$ in three orthogonal directions.
We restricted our study to the subset which have equal contributions
from all directions, controlled by a single parameter $\epsilon$.  Also,
to maintain locality, we restricted ourselves to one step of smearing.

\bef[h]
\begin{center}
\includegraphics[scale=0.40]{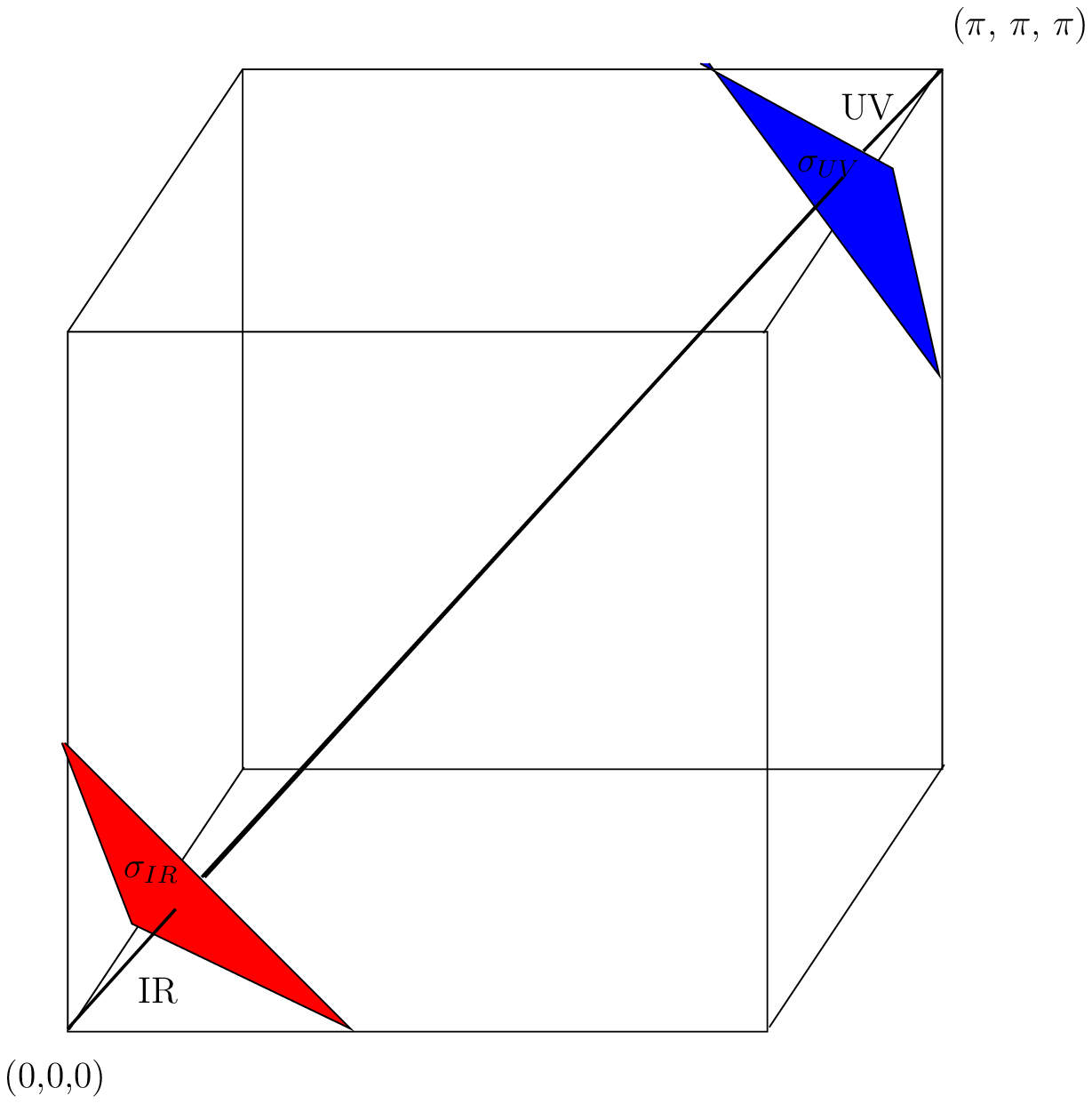}
\caption{
Defining UV and IR. The cube represents a three dimensional Brillouin zone 
for illustration. The body diagonal connects the origin to the opposite 
corner of the cube. The modes between the origin and the plane 
$\sigma_{IR}$, which is perpendicular to the body diagonal, are defined as 
IR modes. Similarly, the modes between $\sigma_{UV}$ and the farthest corner 
are defined as UV. The rest of the modes are generic. The distances of the
planes from the origin are arbitrary.
}
\end{center}
\eef{mode}
To study the effect of gauge link smearing on the high frequency modes of 
the gauge field, we constructed  the power spectrum, $E(k)$, of plaquettes 
summed over all spatial directions, $P(x)$, as 
\beq
 P(k)=\sum_x\exp\left(ik\cdot x\right)P(x) \qquad{\rm and}\qquad
 E(k)=|P(k)|^2.
\eeq{four}
The mode numbers $k_\mu = \pi(2\ell_\mu+\zeta_\mu)/N_\mu$, $N_\mu$
is the size of the lattice in the direction $\mu$, the integers
$0\le\ell_\mu<N_\mu$, and $\zeta_\mu=0$ for periodic boundary conditions
and 1 for anti-periodic. Periodic or anti-periodic boundary conditions
imply that the independent modes are those with $\ell_\mu$ inside the
Brillouin hypercube whose body diagonal, BD, joins the corners (0,0,0,0)
and ($N_x/2,N_y/2,N_z/2,N_t/2$).

We used this power spectrum to find how smearing affects the UV and
IR modes. As shown in \fgn{mode}, we separated the IR and UV using
hyperplanes perpendicular to BD. All modes within the Brillouin zone
closer to the origin than a hyperplane $\sigma_\IR$ were called IR modes;
conversely all modes within the Brillouin zone closer to the far corner
than the plane $\sigma_\UV$ were called UV modes. Everything else was a
generic mode-- neither IR, nor UV.  We defined the suppression of power
in the IR and UV as a function of $\epsilon$
\beq
   Q_\UV=\frac{E_\UV(\epsilon)}{E_\UV(0)}, \qquad{\rm and}\qquad
   Q_\IR=\frac{E_\IR(\epsilon)}{E_\IR(0)},
\eeq{qx}
where $E_\UV(\epsilon)$ is the power summed over all modes in the UV for
a fixed value of $\epsilon$, and $E_\IR(\epsilon)$ is a similar quantity
obtained by summing over all modes in the IR.  
\begin{figure}[]
\begin{center}
\includegraphics[scale=0.35]{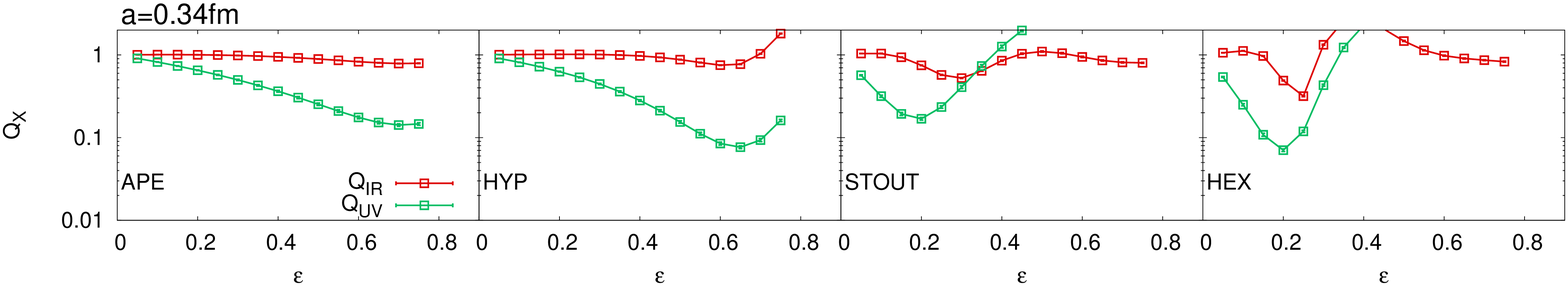}

\includegraphics[scale=0.35]{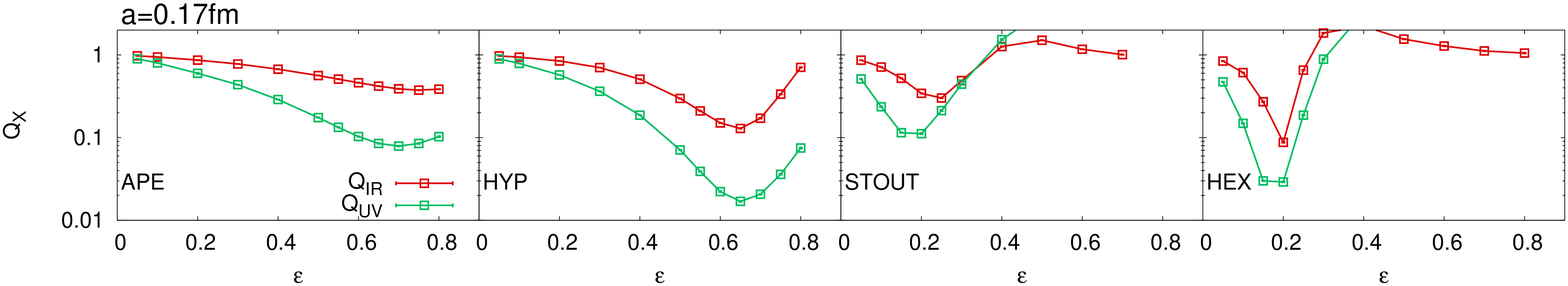}

\includegraphics[scale=0.225]{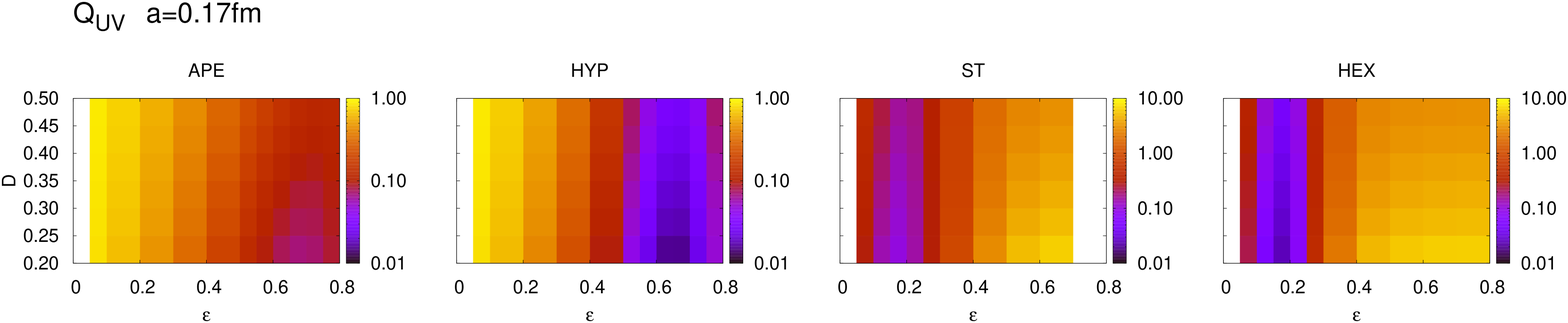}
\caption{\label{fg:Q}Suppression of power. The suppression of power $Q_\IR$
(red points) and $Q_\UV$ (green points) are given as a function of
$\epsilon$ for all the four smearing schemes. The measurements for top
and middle panels are made on configurations with the same pion mass, but
the lattice spacing of the middle panel is half the top one. The bottom
panel shows the variation of $Q_\UV(\epsilon)$ with the distance, $D$,
of $\sigma_\UV$ from  $(\pi,\pi,\pi,\pi)$. The optimum $\epsilon$
remains unaltered.  }
\end{center}
\end{figure}

We investigated $Q$ numerically with thermalized configurations at $T=0$
using $\beta=5.2875$ and $\beta=5.53$ corresponding to lattice spacings
of $0.34$ and $0.17$ fm. The Goldstone pion mass is the same in both
the configurations. Periodic boundary conditions were used so that
all $\zeta_\mu=0$.  The variation of $Q_X$ with $\epsilon$ is shown
in the top two panels of \fgn{Q}. First focusing on the top panel,
one sees that the slope of the curve for $Q_\UV$ always starts off
larger than that for $Q_\IR$. Also, the slope of the latter seems to
be close to zero. This shows that smearing can be used to modify the UV
without modifying the IR.  One can use this to seek an optimum value of
$\epsilon$, such that $Q_\UV$ is as small as possible.  From the bottom
panel, we find that there is change in the overall suppression of power
in the IR and UV, but the change in the optimum $\epsilon$ is not large
even when the lattice spacing is halved. The optimum values of $\epsilon$
move down slightly. This movement is compatible with the intuition that
finer lattices require less improvement.  Since the definitions of IR and
UV are arbitrary, one needs to check whether the results are sensitive
to this definition.  We placed the planes $\sigma_\IR$ and $\sigma_\UV$
at a fraction $D$ of the length of the diagonal (with $0<D<0.5$, so that
no mode is simultaneously in the IR and UV) from the nearest corner,
and varied $D$. The result for $Q_\UV$ is shown in the bottom panel of
\fgn{Q}. The colour code is such that $Q_{UV}$ decreases when we go from
yellow to blue. We find that the optimum $\epsilon$ is insensitive to $D$.
\begin{figure}[]
\begin{center}
\includegraphics[scale=0.35]{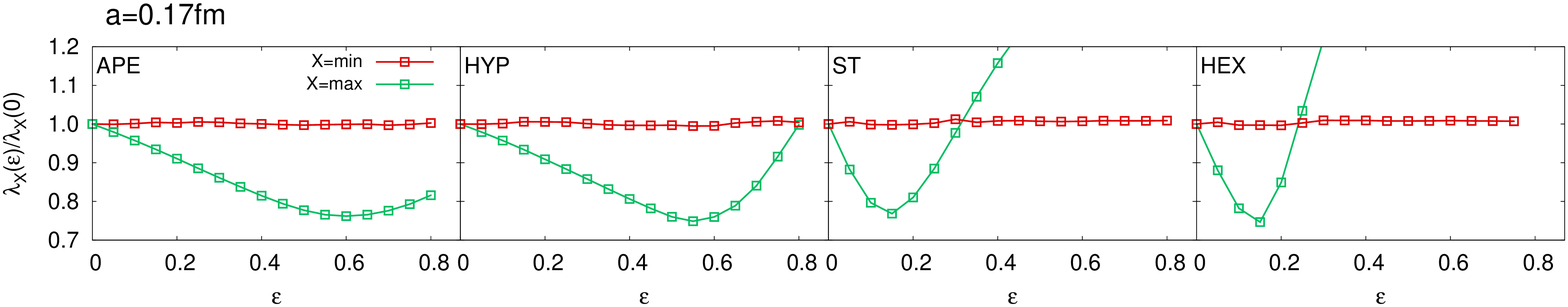}

\includegraphics[scale=0.35]{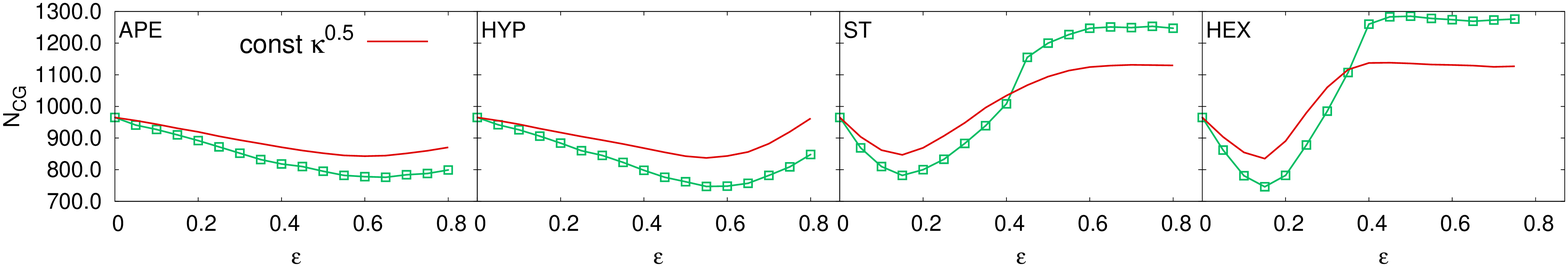}
\caption{\label{fg:lamb}Effect of smearing in quark sector.  The top
panel shows the variation of ratio of extremal eigenvalues (maximum:
green, minimum: red) of $D^\dagger D$ with smearing to their values
without smearing, as a function of $\epsilon$ for all the four smearing
schemes. The green points in bottom panel show the variation of number
of conjugate gradient iterations required for convergence as a function
of $\epsilon$. The red line is a curve proportional to $\sqrt{\kappa}$,
where $\kappa$ is the condition number of $D^\dagger D$. The resemblance
between the two is striking.}
\end{center}
\end{figure}

Then we studied how smearing affected the quark sector. Using
massive staggered Dirac operator, $D$, we took the minimum
and maximum eigenvalues, $\lambda_{min}$ and $\lambda_{max}$,
of $D^\dag D$ to be the IR and UV quantities in the quark
sector. In the top panel of \fgn{lamb}, we show the variation of
$\lambda_{min}(\epsilon)/\lambda_{max}(0)$ with $\epsilon$ by the red
points and similarly for $\lambda_{min}(\epsilon)/\lambda_{max}(0)$
using green points, at $\beta=5.53$. We find that there exists an optimum
$\epsilon$ where $\lambda_{max}$ is minimum. Interestingly, this optimal
point occurs very close to the one determined in the gauge sector. Also,
the changes in $\lambda_{min}$ is very minimal and occurs within the
tolerance used in Lanczos.

The convergence of conjugate gradient (CG) is related to the
extremal eigenvalues through the condition number of $D^\dagger D$,
$\kappa=\lambda_{max}/\lambda_{min}$. In the bottom panel of \fgn{lamb},
we display the variation of number of CG iterations, $N_{CG}$, required
for the residue to be less than $10^{-5}$ as a function of $\epsilon$.
We find that CG is optimized at the same values of $\epsilon$ as found
through $Q_{UV}$ and $\lambda_{max}$. We also show a curve proportional
to $\sqrt{\kappa}$ (red curve) to show the striking dependence of $N_{CG}$
on $\kappa$, even though $\kappa$ only sets an upper bound on $N_{CG}$.

We summarize the results of our optimization of smearing algorithms in 
\tbn{opteps}.

\bet
\begin{center}
\begin{tabular}{|c|ccc|ccc|}
\hline
Scheme 
 & \multicolumn{3}{c|}{$\beta=5.2875$, $am=0.025$}
 & \multicolumn{3}{c|}{$\beta=5.53$, $am=0.0125$} \\
\cline{2-7}
 & $Q_{UV}$ & $N_{CG}$ & $\lambda_{max}$ 
 & $Q_{UV}$ & $N_{CG}$ & $\lambda_{max}$ \\
\hline
APE   & 0.71 & 0.65 & 0.62 & 0.70 & 0.65 & 0.60 \\
HYP   & 0.65 & 0.60 & 0.56 & 0.65 & 0.55 & 0.55 \\
Stout & 0.19 & 0.15 & 0.16 & 0.18 & 0.15 & 0.14 \\
HEX   & 0.20 & 0.15 & 0.17 & 0.17 & 0.15 & 0.14 \\
\hline
\end{tabular}
\end{center}
\caption{The best $\epsilon$ for two different $a$, the second being half
 of the first, evaluated in different schemes and by different optimization
 criteria. The optimum parameter value in each scheme is nearly independent
 of $a$.}
\eet{opteps}

\section{Application to Screening Masses}
We now present the applications of our study on smearing to the study of 
hadronic screening masses. By construction of staggered quarks, each staggered 
meson has 16 different taste partners. In the continuum, the masses of the 
taste partners are degenerate, while at finite $a$, they are split. Hence we 
take the measure of taste breaking as 
\beq
   \delta m_\pi=m_{\gamma_5\gamma_i}-m_{\gamma_5},
\eeq{split}
where the subscript gives the taste $\gamma$ structure. The splitting at
finite temperature, $\delta \mu_{PS}$, is taken as the splitting in the
corresponding screening masses. In \fgn{split}, we show $\delta\mu_{PS}$
at $T=2T_c$ as a function of $\delta m_\pi$. The different points are
labelled by values of $\epsilon$ in different smearing schemes. The
black line is given by $\delta\mu_{PS}\propto (\delta m_\pi)^2$. Thus
the splitting in the deconfined phase improves super-linearly  with the
improvement at zero temperature. This observation could be explained by
complete or almost complete restoration of taste symmetry in the chiral
limit \cite{nik}.  Since recovery of taste symmetry has been used as
the main indicator of the reduction of UV effects, it is natural to use
optimized HYP smearing in order to best reduce lattice artifacts.
\begin{figure}[]
\begin{center}
\includegraphics[scale=0.50]{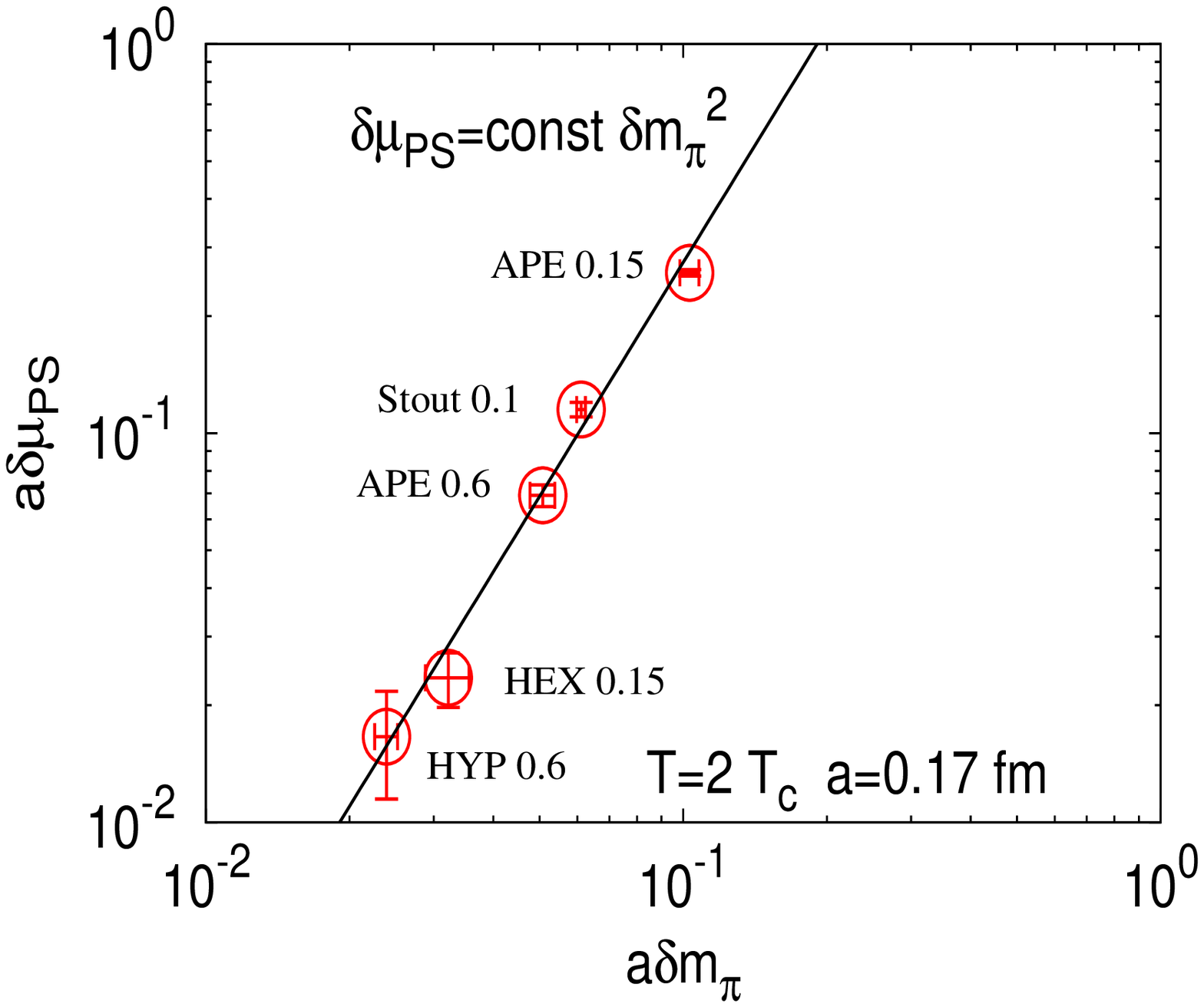}
\caption{\label{fg:split}Super-linear improvement at high $T$. Taste
splittings at $T=0$ and at $T=2T_c$ were determined at the same lattice
spacing $a=0.17$fm. The data points are measured splittings and they are
labelled by the smearing scheme and the value of $\epsilon$ used for its
determination. The black line gives the best fit for $\delta \mu_{PS}
\propto \left(\delta m_\pi\right)^2$.
}
\end{center}
\end{figure}

Using optimal HYP improved staggered valence quark, we determined the
pseudoscalar (PS), scalar (S), vector (V), axialvector (AV) and nucleon
(N) screening masses in the temperature range $0.92 T_c\le T \le 2
T_c$. The results are displayed in \fgn{scrmass} for the ensemble with
$m_\pi \approx 192$ MeV and temporal extent $N_t=4$.
 The blue and green bands are
the weak coupling predictions from dimensional  reduction \cite{dr}
and HTL \cite{htl} respectively. The salient feature of this plot is
that $\mu_{PS}$ in the deconfined phase lies closer to the lattice free
field theory limit. Also, the meson screening masses agree with the weak
coupling theory predictions within the $15\%$ uncertainty arising from
the smearing scheme dependence at this lattice spacing.
\begin{figure}[]
\begin{center}
\includegraphics[scale=0.95]{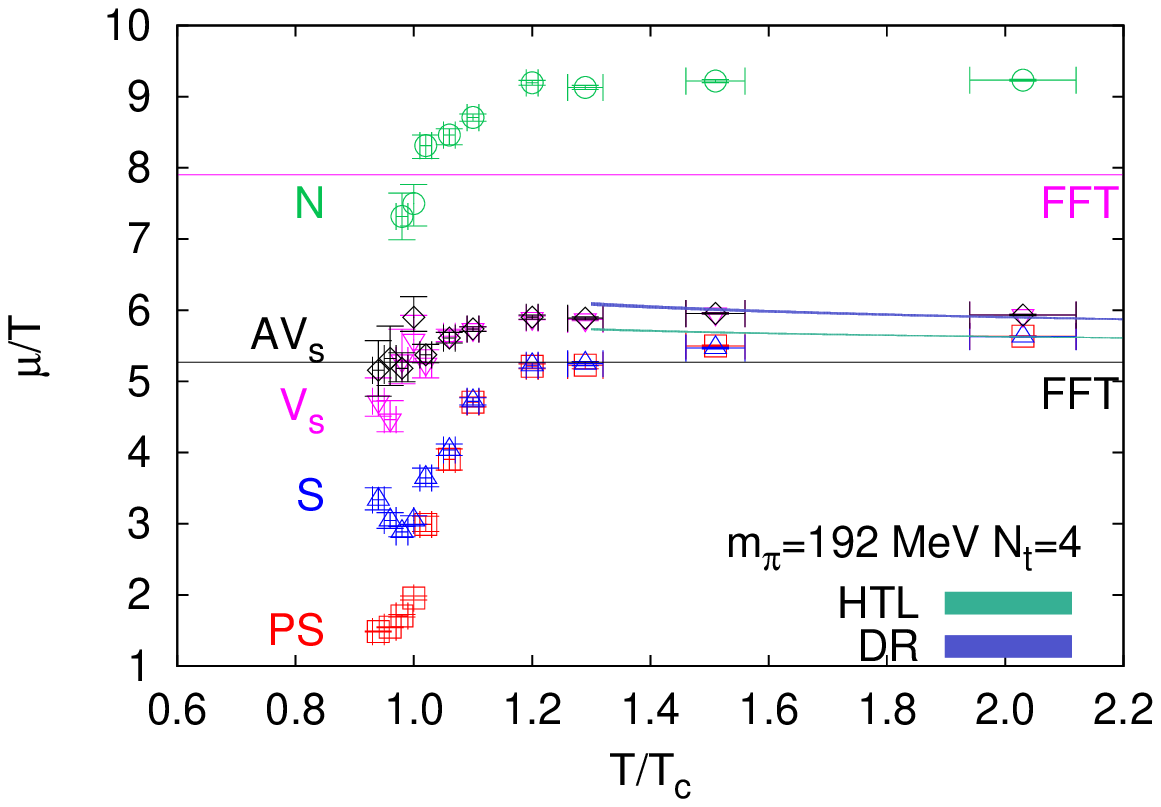}
\caption{\label{fg:scrmass}Hadron screening masses with optimum HYP improved valence quark 
on $N_t=4$ lattice. The blue and the green bands are the weak coupling 
predictions. The solid black line is the lattice free field theory result for 
mesons. }
\end{center}
\end{figure}

\section{Conclusions}
We presented our work on the optimization of the valence quarks.  We used
four popular versions of fat-link staggered quarks.  We optimized the
smearing parameter, $\epsilon$, in each case by observing changes to
the power spectrum of the plaquette (see \fgn{Q}) and the largest and
smallest eigenvalues of the Dirac operator (see \fgn{lamb}). The optimum
$\epsilon$ was chosen so that the UV was suppressed as much as possible
without changing the IR behaviour in both cases.  This also improved the
performance of the conjugate gradient algorithm used for the inversion
of the Dirac operator (see \tbn{opteps}). Such a tuning was done at
$T=0$. We found mild changes in the tuning parameters as the lattice
spacing was changed by a factor of 2.  Smearing causes systematic changes
in finite temperature properties of interest. Taste symmetry breaking
in the hot phase improves super-linearly with improvement at $T=0$
(see \fgn{split}).  Using optimal HYP improved valence quarks,
the screening masses at high temperature are found to be close to
weak-coupling theory (see \fgn{scrmass}).

The lattice computations described here were performed on the Cray
X1 of the ILGTI in TIFR.

\end{document}